%Paper: hep-th/9310153
%From: sadov@string.harvard.edu (Vladimir Sadov)
%Date: Fri, 22 Oct 93 19:35:25 -0400
%Date (revised): Fri, 22 Oct 93 19:43:38 -0400
%Date (revised): Fri, 22 Oct 93 19:53:20 -0400
%Date (revised): Mon, 25 Oct 93 11:25:11 -0400
%Date (revised): Thu, 22 Sep 1994 12:58:20 -0400

\input harvmac
\Title{HUTP-93/A027}{On equivalence of Floer's and quantum
cohomology.}
\centerline{Vladimir Sadov}
\bigskip\centerline{Lyman Laboratory of Physics}
\centerline{Harvard University}\centerline{Cambridge, MA 02138}
\centerline{and}\centerline{L.D. Landau Institute for Theoretical
Physics,
Moscow}

\vskip .3in
We show that the Floer cohomology and quantum cohomology rings of the
almost
K\"ahler manifold $M$, both defined over the Novikov ring of the loop
space
${\cal L}M$, are isomorphic.  We do it using a BRST trivial
deformation of the
topological A-model.  The relevant aspect of noncompactness of the
moduli of
pseudoholomorphic instantons is discussed.  It is shown
nonperturbatively that
any BRST trivial deformation  of A-model which does not change the
dimensions
of BRST cohomology  does not change the topological correlation
functions
either.

\Date{9/93}

\newsec{Introduction}

The ``quantum cohomology"  ring $H_Q^*$ (=(c,c) ring in terms of N=2
sigma models) was introduced in \ref\LeVaWa{W.~Lerche, C.~Vafa,
N.~Warner  {\it
 Chiral Rings in N=2 Superconformal Theory} Nucl.~Phys.~ {\bf B324}
(1989) 427
}, see also \ref\Ca{P.~Candelas, P.~Green, L.~Parkes, X.~de la Ossa
{\it A pair
of Calabi-Yau manifolds as an exactly soluble superconformal field
theory}
Nucl.~Phys.~ {\bf B359} (1991) 21 }, \ref\Va{C.~Vafa {\it Topological
Mirrors
and Quantum Rings} in Essays in Mirror Symmetry, ed. S.~-T.~Yau,
1992},
\ref\WiI{E.~Witten {\it Topological sigma model} Comm.~Math.~Phys.~
{\bf
118}(1988) 411},\ref\WiII{E.~Witten {\it Mirror manifolds and
Topological Field
Theory} in Essays in Mirror Symmetry, ed. S.~-T.~Yau, 1992} and
\ref\AsMo{P.~Aspinwall, P.~D.~Morrison {\it Topological Field Theory
and
Rational Curves} Oxford preprint, 1991}.   The infinite volume limit
of $H^*_Q$
coincides
with the ordinary cohomology ring $H^*(M)$ of the target space $M$.
For  any
finite volume, $H^*_Q$ is a deformation of  $H^*(M)$.    A natural
question
arise about the meaning of this deformation  in classical geometry.

One way to do this in terms of the moduli space of holomorphic
instantons was
introduced in \Ca,\AsMo,\WiII. It is more or less standard by now and
we refer
the reader to \WiII, for a review of that approach. Closely related
to, but not
quite the same as the latter one, is the interpretation in terms of
geometry of
the parameterized loop space ${\cal L}M$ of the target space,
conjectured in
\LeVaWa \Va. It turns out that an appropriate object to deal with in
this
context is what the mathematicians call a Floer symplectic cohomology
$H^*_F$
\ref\FI{A.~Floer {\it Symplectic Fixed Points and Holomorphic
Spheres.}
Comm.~Math.~Phys.~ {\bf 120} (1989) 575-611},
\ref\FII{A.~Floer {\it The unregularised gradient flow of the
symplectic
action.} Comm.~Pure Appl.~Math.~ {\bf 41} (1988)
775-813},\ref\FIII{A.~Floer
{\it Witten's complex and infinite dimensional Morse theory.}
J.~Diff.~Geom.~
{\bf 30} (1989) 207-221}.

 $H^*_F$ appear via the Witten-Floer \ref\Mi{J.~Milnor {\it Lectures
on the
h-cobordism theorem.} Math.~Notes, Princeton Univ.~Press, 1965},
\ref\WiIII{E.~Witten{\it Supersymmetry and Morse theory.}
J.~Diff.~Geom.~ {\bf
17}(1982) 661-692}, \FI complex in ${\cal L}M$, whose vertices are
the fixed
points of some symplectomorphism $\phi$ of $M$ and the edges are the
"pseudoholomorphic instantons" (defined below) connecting these fixed
points.
It is graded by the same abelian group $2\Gamma$ as the quantum
cohomology
$H_Q^*$ (and for the same reason), a phenomenon known to physicists
as the
anomalous conservation of fermionic number. Under some natural
assumptions
\FI,\ref\DoSa{S.~Dostoglou, D.~Salamon {\it Instanton homology and
symplectic
fixed points.} preprint 1992}
one has $dim H^i_F=\sum_{\gamma \in \Gamma}b^{i+2\gamma}(M)$ where on
the left
hand side we identify the index of Betty numbers modulo $2\Gamma$.
Moreover,
there is a natural action of $H^*(M)$ on  $H^*_F$. It is defined in
terms of
intersection numbers in  ${\cal L}M$ of a finite dimensional cycle
--- a cell
of the WF complex --- with a finite codimensional one --- a pullback
of the
cocycle on M under the natural projection  ${\cal L}M \rightarrow M$.
Having
fixed the isomorphisms of vector spaces $h^i: H^i_F \equiv H_Q^i$, we
may think
that we have a new multiplication law (a ring structure) on $H^*(M)$.
This is
quite similar with how it happens for the quantum cohomology ring.

The whole ideology of the Floer theory renders it almost obvious,
that there
should be an isomorphism
\eqn\equival{
H_Q^*\equiv H_F^*
}
Still, there are two obstacles for just the naive identification
\equival.

The first obstacle is that, as in \FI-\FIII, $H^*_F$ is naturally
defined over
integer numbers ${\bf Z}$. In particular, it cannot depend
nontrivialy on any
continuous parameter.
On the other hand, the quantum cohomology $H_Q^*$ is defined over
complex
numbers and its ring structure depends on the K\"ahler structure of
the target
space.

The second problem is that in Floer theory by ``pseudoholomorphic
instantons"
one understands the solutions of the equation
\eqn\pshol{
{{\partial X^i} \over {\partial \tau}}+ J^i_{\  j}{{\partial X^j}
\over
{\partial t}}=\partial ^iH(X,t)
}
where $J^i_{\  j}$ is an almost complex structure on $M$ which
relates the
metrics $G_{ij}$ and the K\"ahler form $k_{ij}$:
\eqn\jgk{
G_{ij}=J^n_{\  i}k_{nj}
}
The function $H$ on the right hand side of the equation (the
Hamiltonian)
depends on the point on $M$ and also  periodic (with period $2\pi$)
on variable $t$. Fix some initial moment $t=0$. The hamiltonian flow,
generated
by $H$, maps $M$ to itself at each $t$. A map, generated when
$t=2\pi$, called
a
period map, gives a symplectomorphism $\phi$. Thus the fixed points
of $\phi$
are in one to one correspondence with the periodic with period $2\pi$
trajectories --- the points of ${\cal L}M$. By difficult analytic
methods
\FI,\FII  Floer has proved that in fact $H_F^*$ is independent of
$H(X,t)$, for
generic $H$. Unfortunately, $H=0$ which gives the usual holomorphic
instantons
equation, is by no means generic for the Morse type theory.

It turns out that the first problem can be dealt with quite easily if
we
redefine both $H_Q^*$ and $H_F^*$ to be defined over some new ring,
called a
Novikov ring \ref\NoI{S.~Novikov {\it Quasiperiodic structures in
topology} in
Topological methods in Mathematical Physics, ed. by L.~Goldberg and
A.~Phillips, Publish or Perish, 1993}, \ref\NoII{S.~Novikov
Russ.~Math.~Surveys
{\bf 37} 5 (1982) 3-49},\ref\ThI{Le Tu Quok Russ.~Math.~Surveys {\bf
44} 3
(1991)}.  This trick is well known to mathematicians
\ref\HoSa{H.~Hofer,
D.~Salamon {\it Floer homology and Novikov rings} preprint
1992},\ref\HoVa{Le
Hong Van, K.~Ono {\it Symplectic fixed points, Calabi invariants and
Novikov
homology} preprint MPI/93-27}\foot{In the last reference the Novikov
ring is
different from ours and used for another reason (to work with
nonexact
symplectomorphisms).}. It also makes possible to work with Floer
theory on the
Calabi-Yau manifold, which otherwise would be impossible.

Our strategy in dealing with the second difficulty will be to show
how the
``pseudoholomorphic instantons" appear in topological sigma model
\WiI,\WiII,\ref\BaSi{L.~Bailieu, I.~Singer {\it The topological sigma
model.}
Comm.~Math.~Phys.~ {\bf 125} (1989) 227}, properly deformed by adding
a
BRST-trivial piece to the action. Thus instead of trying to continue
the Floer
cohomology to the point $H=0$ we extend the quantum cohomology to
arbitrary
$H$ and show that it does not depend on $H$.

We obtain a family of  topological theories  parameterized by the
Hamiltonian
$H$.  For $H=0$  we get back a  topological sigma model of  Witten
\WiI.  The
local physical operators, given by BRST cohomology, are in one-to-one
correspondence with elements of de Rham cohomology of  $M$.
The correlation functions of these operators can be  localized to
holomorphic
instantons. The  operator algebra  is given by  the quantum
cohomology
$H^*_Q$.

For  $H\neq 0$,  the  (off-shell) BRST operator  is the same  as for
$H=0$.
Therefore the  physical operators  are the same.  Moreover, as the
deformation
by $H$ is BRST trivial, the topological correlation functions are the
same.
Hence the operator algebra is  always  $H^*_Q$.   An important
feature of
theories with  $H\neq 0$ is that  it is possible to characterize also
the {\it
states} in a simple fashion. The states of  topological theory are
the ground
(zero energy) states of  the corresponding $N=2$ supersymmetric sigma
model. It
turns out that  perturbatively, these states are in one-to-one
correspondence
with the loops --- critical points of the Floer functional.  The
non-perturbative effects of instantons "lift" some of them, leaving
as the true
ground states only those annihilated by the BRST operator.  The whole
picture
is a  direct  generalization of  one which appears in  Witten's
supersymmetric
quantum mechanics.

Using localization, it is possible to compute the matrix elements of
the BRST
operator between the perturbative ground states.  They turn out to
coincide
with matrix elements of the Floer complex. Therefore the true ground
states
coincide with the elements of  Floer cohomology groups $H^*_F$.  The
operator
algebra  $H^*_Q$ of the theory  acts on the space of states $\cong
H^*_F$. This
action is what we are after.  Using the same localization technique,
one can
find the matrix elements of this representation of   $H^*_Q$ on
$H^*_F$ to be
the same as appear in Floer's theory under the other name.   The
state-operator
 correspondence of topological sigma model leads to identification of
these
matrix elements as  (topological) correlation functions.

This is the outline of both   the idea and  the techniques used in
this paper.
Certainly, it does not contain a {\it mathematicaly}  rigorous proof.
For
example, we take it for granted that  the operator algebra $H^*_Q$ of
topological  sigma model is commutative associative. (Recently this
fact was
proven \ref\RT{Y.~Ruan, G.~Tian {\it Mathematical theory of quantum
cohomology}, preprint  (January, 1994)}.) There are many other fine
points,
part of them discussed in the last section of this paper.  We explain
there why
the BRST trivial deformation which does not change the ranks of  BRST
cohomology, does not  change the topological correlation functions
either.   In
doing this, we cannot use the perturbative argument since we do not
know the
right vacuum of the theory {\it apriory}. The reason is that due to
the
noncompactness of the moduli spaces of pseudoholomorphic instantons,
the state
operator correspondence could furnish a singular map at some points.
The
standard example is the  deformation of  LG model  by the relevant
operator
$W=x^{n+2}\longrightarrow W+\epsilon x^{n+3}$. (In this example, the
BRST
cohomology does change.)  In the last section we show that  such
phenomenon
does not occur in A model, the state operator correspondence is
always smooth
and the correlators are independent of the BRST trivial deformation.

One can understand the relations of the physical approach developed
here with
the  mathematical approach developed by \RT, \ref\Piu{S.~Piunikhin
{\it Quantum
and Floer cohomology have the same ring structure}, preprint
hep-th/9401130} as
follows.  Mathematicians want to  construct a  homotopy of  the Floer
complex
to $H=0$.  It is a technically dificult problem. It turns out that
it is
easier to work with generic {\it almost  complex structure}.  From
the point of
view of  physical approach developed, it is straightforward to
generalize to
that situation.  Since any variation of the almost complex structure
corresponds to BRST trivial deformation of the theory,  the
topological
correlation functions and operator algebra are independent of  it
\WiI,\WiII.
Then, one can interpret   the  constructions of  this paper as a
realization of
  required homotopy on the ground states of  the topological  sigma
model given
by the state-operator correspondence map.  Smoothness of this map is
crucial.

\newsec{Morse theory for multivalued functions: Novikov rings}
In order to organize the material better it seems convenient to begin
with the
explanation of ideas of the Novikov theory before having defined the
Floer
cohomology $H_F^*$ in detail. The only thing we should know about
$H_F^*$ now
is that it appears in a Morse theory for a multivalued function $S$
on the loop
space. That function is defined on the universal cover $\hat{{\cal
L}M}$ of
${\cal L}M$ (it is necessarily abelian) and changes by $\oint_\gamma
k$ as one
moves along $\hat{{\cal L}M}$ by $\gamma \in \pi _1 ({\cal L}M)=\pi
_2(M)$, $k$
is a K\"ahler form.

Consider first a situation in general, following closely the
presentation of
\NoI. Let $X$ be a closed manifold. We do not specify whether it is
finite
dimensional or not. Of course, in the latter case, which appears in
Floer's
theory we have problems with compactness, but here we want to forget
about it
for a moment. Let $\gamma _1,\ldots ,\gamma _n$ be a basis for the
first
homology group of X. For every closed 1-form $\omega$ on X we have n
periods
\eqn\per{
k_j=\oint_{\gamma _j}\omega
}
The numbers $k_1,\ldots ,k_n$ are in general irrational and their
linear
combinations with integral coefficients form a free
 abelian group. The rank $k$ of this group is called the {\it
irrationality} of
1-form $\omega$. Obviously, $k\leq n$. From now on we suppose that
$k=n$ which
means that $\omega$ is "generic enough".

There is a minimal free abelian covering $p:\ \bar{X}\rightarrow X$
such that
the form $p^*\omega$ is exact:
\eqn\exec{
p^*\omega=dS
}
The monodromy group is ${\bf Z^n}\equiv H_1(X)$, generated by the
covering
transformations $T_i:\ \bar{X}\rightarrow \bar{X}$, satisfying
$T^*_iS=S+k_i$.
Take on $X$ a smooth metric such that the hamiltonian flow generated
by
$\omega$ lifts smoothly to a $\infty$-continuous flow on the covering
space
$\bar{X}$: each trajectory ends in a critical point or intersects all
the
level-surfaces of the function $S$ on $\bar{X}$. Consider now a
cellular
decomposition $\cal C$ (with the structure of complex) of $X$. For
example, it
can be a Morse decomposition defined as a collection of the surfaces
of
steepest descent starting from the critical points.
This gives a collection of cells
\eqn\cells{
\sigma ^i_q,\ q=1,\ldots ,m_i
}
The complex $\cal C$ lifts to a complex $\hat{\cal C}$ in $\bar{X}$
with a free
action of the monodromy group ${\bf Z^n}$ on it. We can denote the
cells of
$\hat{\cal C}$ by
\eqn\cellc{
t_1^{s_1}t_2^{s_2}\cdots t_n^{s_n}\sigma ^i_q,\ q=1,\ldots ,m_i
}
then the generators $T_i^{\pm 1}$ of the monodromy group act by
multiplication
by $t_i^{\pm 1}$. We define the boundary operator on $\hat{\cal C}$
by
\eqn\bouI{
\partial \sigma ^i_q=\sum_p a_{pq}(t_1,\ldots ,t_n)\sigma _p^{i-1}
}
where the coefficients $a_{pq}(t_1,\ldots ,t_n)$ are the formal
series in $t_i,
t_i^{-1}$. To be precise, let us give a definition

{\it Definition} A ring $K_n$ (a Novikov ring) consists of all such
formal
power series in $t_1,\ldots ,t_n,t_1^{-1},\ldots ,t_n^{-1}$ that the
following
two conditions are met:

a)\  There exists a number $N(a)$ such that if
\eqn\condI{
a=\sum u_{s_1,\ldots ,s_n}t_1^{s_1}\cdots t_n^{s_n}
}
then the coefficient $u_{s_1,\ldots ,s_n}=0$ if $\sum_j s_jk_j<N(a)$.

b)\ There is only a finite number of nonzero coefficients in any
domain
\eqn\condII{
N_1<\sum_j s_jk_j<N_2
}

{\it Example} If n=1, then
\eqn\exnI{
K_n=Z[t^{-1},t]]\equiv \bigl\{a=\sum_{n\geq N(a)}u_nt^n,\ u_n\in
Z\bigr\}
}
--- the Novikov ring coincides with all formal series with the finite
negative
part.

There is a natural embedding of the group ring of the monodromy group
${\bf
Z^k}$ to $K_n$:
\eqn\locs{
0\rightarrow Z[t_1,\ldots ,t_n,t_1^{-1},\ldots ,t_n^{-1}]\rightarrow
K_n
}
This embedding generates the local system $\cal K$ on $X$ with
coefficients in
the ring $K_n$ and the corresponding homology groups $H_*(X,{\cal
K})$ are
$K_n$-modules. It is the boundary operator of this local system
complex that we
have written in \bouI.

Having established the basic facts about Novikov rings in general
situations,
we are back to the loop spaces ${\cal L}M$.
We have the canonical isomorphism\foot{
We suppose that $M$ is simply connected.}
\eqn\trivid{
H_1({\cal L}M)=H_2(M)=\pi _2(M)
}
To define  a  basic 1-form  $\omega$ on the loop space  ${\cal L}M$
let us
notice that the value of  vector field $\xi (z) \in Vect({\cal L}M)$
in the point $z=\left\{ X(t),\;0\leq t\leq 2 \pi \right\} \in {\cal
L}M $   is
a vector field $ \xi (t)$ on the circle  $ z\subset M$. Then the
value of  the
form $ \omega$ on $ \xi$
in $z \in {\cal L}M$ equals
\eqn\form{
\omega \left( \xi \right) \left( z \right) = \oint_{z}{k(\xi  ( t )),
}
}
where $k$ is a {\it K\"ahler}  2-form\foot{That is, $[k]\in H^2(M)$
is positive
on all the pseudoholomorphic curves in $M$. } on $M$.
Now let us apply what we have developed above to the quantum
cohomology ring.
One notices immediately that $H_Q^*$ is in fact already defined over
$K_n$, if
one identifies
\eqn\ident{
t_i=e^{\oint_{\gamma _i}k}
}
Indeed, both the fundamental two- and three-forms, defining $H_Q^*$
take values
in series in $t_1,\ldots ,t_n$ with integer coefficients and easy to
see that
actually they take values in $K_n$. Moreover, we know that there can
only
appear the {\it positive} degrees of the generators $t_1,\ldots ,t_n$
in all
these formulas. This is because we only consider the {\it
holomorphic} maps
into M for such maps the degree is always nonnegative. Thus in the
definition
of the Novikov ring given above, we can restrict ourselves to the
series having
no negative part. With this remark, we will use the term "Novikov
ring" meaning
the ring for the loop space constructed above.

{}From this point of view on the quantum cohomology we think of the
variables
$t_1, \ldots ,t_n$ as of indeterminates, but define $H_Q^*$ over a
bigger ring
$K_n$ instead of complex numbers $C$, so the Betti numbers do not
change
(generically).

\newsec{A brief review of Floer symplectic cohomology}
The purpose of this section is to briefly discuss the Floer theory in
a way
clarifying its resemblance to the quantum cohomology (modulo two
obstacles,
mentioned in the introduction). For example, from the very beginning
we define
it over the Novikov ring. Also, we don't try to give any proofs in
this
section, referring the reader to
\FI-\FIII,\DoSa,\HoSa,\HoVa,\ref\Sa{D.~Salamon
{\it Morse theory, the Conley index and the Floer homology}
Bull.~L.~M.~S.~
{\bf 22} (1990)113-140},\ref\SaZe{D.~Salamon, E.~Zehnder {\it Morse
theory for
periodic solutions of Hamiltonian equations and the Maslov index}
Comm.~Pure
Appl.~Math.~ {\rm XLV} (1992) 1303-1360}.

Let $M$ be a K\"ahler manifold with a  K\"ahler form $k$. This closed
2-form
defines a symplectic structure on $M$, providing a one to one map
between
vector fields $v$ and 1-forms $\omega$ on $M$ given by the formula
$\omega =k(v,\cdot )$. A vector field $v$ preserves $k$ iff $\omega$
is closed;
$v$ is called {\it hamiltonian} iff $\omega =dH$ is exact. A function
$H$ is
called a Hamiltonian.

The Hamiltonian equation
\eqn\Hame{
{dX^i \over dt}=v^i(X,t)
}
defines a family $u_H(t)$ of diffeomorphisms of $M$, preserving $k$
(called
symplectomorphisms). They are  characterized by the condition that
$X^i(t)=u^i_H(t,X)$ solves \Hame for all $X \in M$.
The Floer theory studies fixed points of the period map
$Per:\;u_H(t,X)
\rightarrow u_H(t+2\pi ,X)$ for the Hamiltonian flows with periodic
in $t$
hamiltonians:
\eqn\peri{
H(X,t+2\pi )=H(X,t)
}
Such a points are in a one to one correspondence with the periodic
trajectories
of \Hame having a period exactly $2\pi$.
It coincides with the ordinary Morse theory on $M$ when $H$ is
independent of
$t$, as the fixed points of $u_H(t)$ are just the critical points of
$H(X)$
then. For the time dependant hamiltonians $H(X,t)$ the Floer theory
is a sort
of a Morse theory on the loop space ${\cal L}M$ of $M$.

Let us consider a (multivalued) function $S_H$ on ${\cal L}M$
\eqn\Morf{
S_H(z)=\int_{D^2}\phi ^*(k)+\int_{S^1}H(X(t),t)dt
}
Here $z$ is a point in ${\cal L}M$
\eqn\loopo{
z=\bigl\{X(t)|X(0)=X(2\pi )\bigr\}
}
and a smooth function $\phi$ furnishes a map of a disk
$\phi:D^2\rightarrow M$
with the boundary values $X(t)$, so $\partial D^2=S^1$. Since $dk=0$,
$S_H$
depends only on the homotopic type of $\phi$ with fixed boundary.
This function
becomes single valued on the minimal abelian cover $\widehat{{\cal
L}M}$ of
${\cal L}M$ (with monodromy group ${\bf Z^k}$ where $k=rankH^2(M)$).
When $\pi
_2(M)$ has no torsion, $\widehat{{\cal L}M}$ coincides with the
universal cover
of ${\cal L}M$.

For a smooth vector field $\xi$ on ${\cal L}M$ (in the  point $z \in
{\cal L}M$
it gives a vector field over the contour $z$ in $M$) the
$\xi$-derivative of
$S_H$ is well definite:
\eqn\derS{
(\xi \cdot D)S_H(z)=\oint_{S^1}\{k(z'(t),\xi (t))+dH(\xi
(t))(X(t),t)\}dt
}
A point $z$ is a critical point of $S_H$ iff \derS vanishes for all
$\xi$,
which happens iff $z=X(t)$ satisfies \Hame. As $z \in {\cal L}M$ is
periodic
with the period $2\pi $ by definition, it gives the fixed point of
the period
map we are after.

The trajectories of the gradient flow of $S_H$ are the
solutions\foot{They are
also called the pseudoholomorphic instantons.} $X^i(t,\tau ):
S^1\times
R\rightarrow M$ of the partial differential equation
\pshol. Two terms
\eqn\veloo{
G^i(X(t,\tau ))=J^i_{\  j}{{\partial X^j} \over {\partial
t}}-\partial ^iH(X,t)
}
may be considered as a vector field on ${\cal L}M$ evaluated at
$z(\tau )$ so
\pshol is a gradient flow equation on ${\cal L}M$:
\eqn\grafl{
{{\partial z(\tau )}\over {\partial \tau }}=-G(z)
}
On the other hand, when $H(X,t)=0$ \pshol is just the Cauchy-Riemann
equation
for the holomorphic instantons.

The function $S_H(z)$ decreases along the trajectories of \grafl
\eqn\decr{
{{\partial S_H(z)}\over {\partial \tau}}=-\oint_{S^1}\Bigl|{{\partial
X(t,\tau
)} \over {\partial \tau}}\Bigr|^2dt
}
Thus the set of trajectories for which
$\int_R\int_{S^1}\bigl|{{\partial
X(t,\tau )} \over {\partial \tau}}\bigr|^2dtd\tau$ is finite
coincides with one
of those for which $S_H$ is bounded. Such trajectories connect the
critical
points of $S_H$.
We define the Morse complex ${\cal C}_H$ as the set of bounded
trajectories
\eqn\Morsc{
{\cal C}_H=\bigl\{X(t,\tau)-{\rm \; a\; solution\; of\;
\pshol\;}\bigl|
\int_R\int_{S^1}\Bigl|{{\partial X(t,\tau )} \over {\partial
\tau}}\Bigr|^2dtd\tau <\infty \Bigr\}
}
Let us define ${\cal C}_H(z_+,z_-)$ as a set of trajectories in
${\cal C}_H$
such that $z(\tau )\rightarrow z_\pm$ when $\tau \rightarrow \pm
\infty$ and a
set
of $k$-trajectories ${\cal C}_H^k(x,y)$ going from $x$ to $y$ as the
set of all
$k$-tuples $z_1(\tau ),\ldots ,z_k(\tau )$ such that $z_i(\tau )\in
{\cal
C}_H(x_{i-1},x_i)$, $x_0=x$, $x_k=y$. A shift of the variable $\tau$
preserves
${\cal C}_H(z_+,z_-)$ so it makes sense to consider the quotient by
the
translational symmetry
\eqn\chat{
{{\widehat{{\cal C}}}_H(z_+,z_-)}={\cal C}_H(z_+,z_-)/{\bf R}
}

The set $\cal Z$ of the critical points is graded by the analog of
the Morse
index $\mu$. But unless $c_1(M)$=0, i.e. unless $M$ is a Calabi-Yau
manifold,
this is not a ${\bf Z}$ grading. Let $\Gamma \subset Z$ be a lattice
generated
by the
set of periods of $c_1(M)$ on $\pi _2(M)$, then there is a function
$\mu :{\cal
Z}\rightarrow Z/{2\Gamma}$ such that
\eqn\dimec{
dim{\cal C}_H(x,y)=[\mu (x)-\mu (y)]\; (mod\; 2\Gamma )
}
There is the same situation for the quantum cohomology, where $\mu$
is called a
fermionic number. The ambiguity in \dimec occurs because
\ref\Gr{M.~Gromov {\it
Pseudoholomorphic curves in Symplectic Manifolds}, Invent.~Math.~
{\bf 82}
(1985) 307-347},\FI a sequence of paths in ${\cal L}M$ can diverge by
splitting
off a (pseudo)holomorphic sphere (instanton) $w$, and for a joint of
a path
$z(\tau )\in {\cal C}_H(x,y)$ with $w$ we have
\eqn\anom{
\mu (z\# w)=\mu (x)-\mu (y)+2c_1(w)
}
This "splitting off a sphere" phenomenon also results in that the
compactness
properties of the cells ${{\widehat{{\cal C}}}_H(z_+,z_-)}$ are not
so good as
they are in the finite dimensional Morse theory. But it can happen
only for
those components with dimensions bigger then 2. ( Basically, because
an $S^1$
action on the 2-sphere gives an additional degree of freedom.) Thus,
as it were
in finite dimensional case, the 0-dimensional component of
${{\widehat{{\cal
C}}}_H(z_+,z_-)}$ is finite and the 1-dimensional component is
compact up to
the boundaries from ${{\widehat{{\cal C}}}_H^2(z_+,z_-)}$.

Now let $ Z^*$ be a free module generated by the  critical points
$\cal Z$ over
the Novikov ring $K$. This module is graded by $\mu$.
For every isolated trajectory $z(\tau )$ belonging to the
0-dimensional
component of ${{\widehat{{\cal C}}}_H(z_+,z_-)}$, let $\sigma (z)$
denote its
orientation Wi, F and $\rho (z)=t_1^{s_1(z)}\cdots t_n^{s_n(z)}$
denote the
homomorphism which the local system $\cal K$ (defined in sec.2)
associates with
the path $z(\tau )$. We define the matrix element of the coboundary
operator by
the formula
\eqn\cobou{
<\delta y,x>=\sum_z \sigma (z)\rho (z)
}
where the sum is taken over all isolated trajectories in ${\cal
C}_H(z_+,z_-)$.
Then the action of the coboundary operator on $Z^*$ is given by
\eqn\coboa{
\delta y=\sum_{x\in {\cal Z}}<\delta y,x>x
}
These formulas are to be compared with that obtained in \WiIII for
the finite
dimensional situation:
\eqn\Wibo{
<\delta y,x>=\sum_z \sigma (z)e^{h(y)-h(x)}
}
When the Morse function is single valued, as in \WiIII, the factors
$e^h(x)$
can be absorbed by redefinition of the vertices $x$. When this is not
the case,
as in the Floer theory, the set of "phase factors" $\rho
(z)=e^{h(y)-h(x)}$
forms the nontrivial 1-cocycle on $\cal Z$ and cannot be canceled out
by any
renormalization. We see that the formula \Wibo, which naturally
appears in the
context of supersymmetric quantum mechanics, knows already about the
Novikov
ring. The formula \cobou should be considered as its counterpart for
the
topological sigma model where it computes the matrix element of the
BRST
operator between two wavefunctionals localized on the loops $x$ and
$y$
respectively.

Computing the intersection numbers ( weighed by $\rho$ ) of cycles in
${\cal
L}M$ with cells ${\cal C}_H(z_+,z_-)$ of the Floer complex, we could
define the
cup product $\cup :H^*({\cal L}M)\times H^*_F\rightarrow H^*_F$.
But because of noncompactness the intersection numbers are only
defined for the
particular dual classes of $H^*({\cal L}M)$, pulled back from $M$ by
the zero
time evaluation map $\pi :z=\{X(t)|X(0)=X(2\pi )\}\rightarrow X(0)$.
This is
similar to what we have for the quantum cohomology where we should
compute the
intersections only those cycles on the moduli space pulled back from
$M$.

In order to define a restricted cup operation $\alpha \cup :
H^*_F\rightarrow
H^*_F$ we represent the cohomology class $\alpha \in H^p(M)$ by the
dual
singular simplex $\alpha:$ $\bigcup_i \Delta _i^{|\alpha
|}\rightarrow M$,
$|\alpha |=dimM-p$ and set
\eqn\inter{
\alpha \cap {\cal C}_H(x,y)=\bigl\{(\lambda ,z)\in \cup \Delta
^{|\alpha
|}\times {\cal C}_H(x,y)\bigl| \alpha (\lambda )=X(0)\bigr\}
}
We have $dim(\alpha \cap {\cal C}_H(x,y))=p-dimM$. Then we define the
weighed
intersection number as
\eqn\wein{
<x, \alpha \cup y>=\sum_p \sigma (p)\rho (p)
}
where $p$ runs over 0-dimensional part of $dim(\alpha \cap {\cal
C}_H(x,y))$
and $\sigma (p)$ is the usual relative orientation factor $\pm 1$.
The point
$p$ lies on one particular path $z_p(\tau )$ and
\eqn\weght{
\rho (p)=\rho (z_p)=t_1^{s_1}\cdots t^{s_n}_n
}
 is the homomorphism which the local system $\cal K$ associates to
$z_p$.
Finally, the cup operation $\alpha \cup : Z^*\rightarrow  Z^*$ is
defined as
\eqn\cupo{
\alpha \cup y=\sum_{x\in {\cal Z}}x<x, \alpha \cup y>
}
So defined the cup operation commutes with the coboundary operator
$\delta$ and
therefore descends to $H^*_F$. The subtlety here is that for two
arbitrary
$x,y\in Z^*$ the matrix element \wein depends on $\alpha$ itself, not
only on
its cohomology class\foot{ I thank D.~Kazhdan who pointed out this
fact.}. It
becomes independent of the choice of any particular representative
for $[\alpha
]\in H^*(M)$ only after we descent from $Z^*$ to $H_F^*$. We will
understand
this better in the next section, in terms of decoupling of
BRST-trivial states
from the correlation functions of topological sigma model.

{}From the point of view of the topological sigma model the bracket
$<x, \alpha
\cup y>\equiv <x|\alpha |y>$ should be interpreted as a matrix
element of the
operators corresponding to $\alpha$, taken between vacua $x$ and $y$.
In the next section we give such interpretation and relate these
matrix
elements with  three point correlation functions of the A-model.

In the Floer theory, we consider the elements of de Rham cohomology
of $M$ as
linear operators, acting on $H_F^*$, so there is a homomorphism
\eqn\homo{
\nu :H_{deRham}^*(M) \longrightarrow End(H^*_F)
}
It is not obvious, that the image of $\nu$ is a {\it ring}, i.e. that
it is
preserved by the operator multiplication\foot{I thank I.~Singer and
C.~Taubes
for the discussions that helped to realize importance of this.}. We
shall see
it is true only when we identify this image as the operator algebra
of
topological A-model, which is closed  (and associative \RT). It would
be very
interesting to be able
to prove this fact directly from the definitions \wein, \cupo.

\newsec{Floer theory as a topological quantum field theory}
\subsec{Pseudoholomorphic instantons in the topological A-model}
In this section we give a physical interpretation of the Floer theory
in
terms of the topological sigma model \WiI,\BaSi,\WiII. It leads to
identification of $H_F^*$ as a quantum cohomology ring.

As usual, we start from the $N=2$ supersymmetric sigma model and
perform a
topological twist so that one of the SUSY generators becomes a
1-form. Then the
corresponding charge is the BRST operator $Q$. The $N=2$ chiral
multiplet of
fields of the model contains
\eqn\contab{\eqalign{
{\rm Bosons:\;} & {\rm world\; sheet\; scalar\;} X^i\; - {\rm
target\; space\;
coordinates\;} \cr
        & {\rm world\; sheet\; one\; form\;} F_\alpha ^i - {\rm
target\;
space\; vector\;} \cr
{\rm Fermions:\;} & {\rm world\; sheet\; scalar\;} \chi ^i - {\rm
target\;
space\; vector\;} \cr
          & {\rm world\; sheet\; one\; form\;} \rho _\alpha ^i - {\rm
target\;
space\; vector.} \cr
}}
The field $F_\alpha ^i$ is what is called the auxiliary field. Both
$\rho
_\alpha ^i$ and $F_\alpha ^i$ satisfy a self duality constraint
\eqn\sfdu{\eqalign{
&\rho ^i_\alpha =i\epsilon _\alpha ^{\ \beta}J^i_{\ j}\rho ^j_\beta
\cr
&F^i_\alpha =\epsilon _\alpha ^{\ \beta}J^i_{\ j}F^j_\beta \cr
}}
 The name 'auxillary' stresses that $F_\alpha ^i$ serves to close
$N=2$ algebra
off shell and that it can be set to zero on shell. Here we want to
show how it
can be used to localize the path integral of the topological theory
to the
pseudoholomorphic instantons satisfying \pshol with a nontrivial
right hand
side.

The BRST action on the fields of the multiplet is given by
\eqn\brst{\eqalign{
&[Q,X^i]=i\chi ^i \cr
&\{Q,\chi ^i\}=0 \cr
&\{Q,\rho ^i_\alpha \}=F_\alpha ^i+\partial _\alpha X^i+\epsilon ^{\
\beta}_\alpha  J^i_{\ j}\partial _\beta X^j -i\Gamma ^i_{jk}\chi
^j\rho
^k_\alpha +{i\over 2}\epsilon ^{\ \beta}_\alpha D_k J^i_{\ j}\chi
^k\rho
^j_\beta \cr
}}
We don't need an awkward explicit formula for the commutator
$[Q,F_\alpha ^i]$,
it is enough to know that it is fermionic and equals to zero on the
subvariety
$\chi ^i=0, \rho ^i_\alpha =0$ in the field space.

The local physical operators (observables) of the model are the BRST
cohomology,
isomorphic to de Rham cohomology of $M$ \WiI,\BaSi. To any $p$-form
$\omega
=A_{i_1\cdots i_p}dx^{i_1}\cdots dx^{i_p}$ there corresponds an
operator\foot{
This local operator
is a scalar on the world sheet. Besides, there are the nonlocal
physical
operators which
are integrals of 1- and 2-forms, the whole hierarchy related by the
"descent
equation"
\WiII.} ${\cal
O}_\omega = A_{i_1\cdots i_p}\chi ^{i_1}\cdots \chi ^{i_p}$ and
\eqn\brac{
[Q,{\cal O}_\omega ]={\cal O}_{d\omega }
}
The isomorphism $H^*_{BRST} = H^*_{de Rham}(M)$ follows from \brac.
It is
important to note that  we computed the {\it off-shell}  BRST
cohomology.  We
see that the auxiliary field  $F_\alpha ^i$ does not  show up in the
formula
for  ${\cal O}_\omega$. (It does appear   in the nonlocal  physical
operators
though. But in this paper we are only concerned with the local
operators, which
by the state operator correspondence are related to the ground
states. )

The next step is the computation of the matrix elements of the
physical
operators (the correlators). First of all, we should specify our two
dimensional action. The standard choice, coming from $N=2$ sigma
model, is
\eqn\sact{
A_0=\int X^*(k)+\bigl\{Q,\int{1 \over 2}g^{\alpha \beta}G_{ij}\rho
^i_\alpha
(\partial _\alpha X^j-{1\over 2}F^j_\beta )\bigr\}
}
Then using the BRST fermionic symmetry, the path integral
\eqn\FIT{
<{\cal O}_1\cdots {\cal O}_m>=\int {\cal D}X^i{\cal D}\chi ^i{\cal
D}\rho
^i_\alpha {\cal D}F ^i_\alpha \;{\cal O}_1\cdots {\cal O}_m
e^{A_0[X,\chi ,\rho
,F]}
}
can be localized \WiII on the variety ${\cal E}_0$ of fixed points of
$Q$
(classically, we should treat $Q$ as a vector field in the field
space).

But first we want to get rid off the auxiliary field $F^i_\alpha$. It
can be
done either by taking the Gaussian in $F^i_\alpha$ integral \FIT or
by using
the equations of motion for $F^i_\alpha$.
If we choose the standard  action \sact, the (algebraic) equations of
motion
give $F^i_\alpha =0$ so in \FIT we can drop simultaneously the
integration in
$F^i_\alpha$ and $F^i_\alpha$-dependent piece of the action in the
exponent.

{}From \brst we see that on ${\cal E}_0$ the fermions vanish: $\chi
^i=0, \rho
^i_\alpha =0$ and the fields $X^i(z,\bar{z})$ satisfy the
Cauchy-Riemann
equation for the holomorphic maps from the world sheet into $M$.

Up to now this was a well known story \WiII\ about A-model, leading
to the
notion of the quantum cohomology ring, characterized by the two- and
three-
point correlation functions on sphere. At this moment we can note
that the
auxiliary field $F^i_\alpha$ is bosonic. Hence there is nothing wrong
if it has
a
nontrivial expectation $\Phi ^i_\alpha(X,z,\bar{z})$ on shell.  The
expectation
$\Phi
^i_\alpha(X,z,\bar{z})$ should satisfy the same self duality
condition \sfdu as
$F^i_\alpha$.
We can {\it actually} give to $F^i_\alpha$ the expectation $\Phi
^i_\alpha(X)$
on shell,
if we add to the action $A_0$ a BRST trivial piece
\eqn\defo{
A=A_0+\bigl\{Q,\int g^{\alpha \beta}\rho _\alpha ^i\Phi
^i_{\beta}\bigr\}
}
In the sense of topological theory, the deformation \defo\ is
trivial.  We
explicitely
see that  the  {\it local} physical operators are independent of
$\Phi^i_\alpha$, since {\it off shell} they do not  contain
$F^i_\alpha$.
Also,  all
the correlation functions \FIT\ remain the same.  This last point is
not quite
trivial because of  the possible problems with noncompactness. In
Section 6 we
argue that these problems do not really appear in our setup.

The conditions \brst\ gives now for the $X^i$ fields on the fixed
points locus
${\cal E}_H$ the equation
\eqn\psholI{
\partial _\alpha X^i+\epsilon ^{\ \beta}_\alpha  J^i_{\ j}\partial
_\beta X^j=
-\Phi ^i_\alpha
}
(and the fermions vanish on ${\cal E}_H$ as before).

To study the Floer theory, we only need the case when the worldsheet
is a
cylinder $S^1\times R^1$. Then we can introduce the global
coordinates $(t\in
S^1,\ \tau \in R^1)$, the same as in section 3. The field $\Phi
^i_\alpha$ has
two components, related to each other by \sfdu:
\eqn\reduc{
\Phi ^i_t=J^i_{\ j}\Phi ^j_\tau
}
On the cylinder we can consistently take $\Phi ^i_\alpha (X,t)$ to be
a
(periodic) function of the space coordinate $t$ independent of the
time
coordinate $\tau$. Then the energy is conserved.

In the section 4.2 we shall show that in order to have a reasonable
fermionic
sector we need to consider only the {\it hamiltonian} vector fields
$\Phi
^j_\tau$, that is
\eqn\haml{
\Phi ^j_\tau =-\partial ^jH
}
The function $H(X,t)$ is a hamiltonian. Then \psholI\ is just the
pseudoholomorphic instantons equation \pshol we know from the Floer
theory.

The only geometry of the world sheet we need to consider to compute
the matrix
elements \cobou, \wein in the Floer theory is that of the cylinder.
Suppose now
that our A-model lives over an arbitrary Riemann surface. Let us
briefly
discuss what are the restrictions on $\Phi ^i_\alpha (X,z,\bar{z})$.
First,
there is no good way to divide globally the coordinates into space
and time. If
we want, as we usually do, that the energy be conserved we have to
consider
only the coordinate independent fields $\Phi ^i_\alpha (X)$. Second,
the
consistency of the fermionic sector now requires that the {\it both}
components
of $\Phi ^i_\alpha (X)$ be the hamiltonian fields. This condition,
together
with selfduality \sfdu, forms an overdetermined system of equations
for two
hamiltonians $H_\alpha (X)$. The compatibility condition for this
system
equivalent to requirement for $J^i_{\ j}$ to be {\it a complex
structure} on
$M$.

\subsec{The operators and the states}
The family ${\cal E}_H$ of deformations of the standard variety
${\cal E}_0$
gives a family of localizations of the same topological field theory.
Hence the
set of the observables for ${\cal E}_H$ is the same as for ${\cal
E}_0$ and
coincides with the de Rham cohomology $H^*_{de Rham}(M)$. What is new
is that
for generic $H\neq 0$ it is possible  to localize the {\it states} in
the
theory to the critical loops set $\cal Z$ of the Floer complex.
Indeed, the action is (we work on the cylinder):
\eqn\redac{
A=\int \bigl\{(\partial _\tau X^i+J^i_{\ j}\partial _tX^j+\Phi
^i_\tau )^2
+ig^{\alpha \beta}\rho ^i_\alpha D_\beta \chi ^i + ... \bigr\}dtd\tau
}
so the bosonic piece of the potential energy of the string
configuration $z$ is
 given by
\eqn\poten{
E[z]=\oint\bigl(J^i_{\ j}{{dX^i} \over {dt}}+\Phi ^i_\tau \bigr)^2dt
}
The set of the minima of $E[z]$ coincides with $\cal Z$. Note that
the
potential energy functional comes entirely from the BRST trivial
piece of the
action. Making the coefficient before it arbitrarily large we make
the walls
around minima arbitrarily steep, so classically the string never
flies away
from the minima. Thus the wavefunctionals $\Psi$ of the physical
states should
be the linear combinations of those $|x_i>$ localized to $x_i\in
{\cal Z}$
\eqn\physt{
\Psi =\sum_{x_i\in {\cal Z}} \lambda _i |x_i>
}
additionally satisfying the BRST condition $Q\Psi =0$, modulo
$Q$-trivial
vectors.

The quadratic in fermions piece of the action is (we have used the
equations of
motion for $X^i$ to simplify it)
\eqn\feac{
A_F=\int \bigl\{-iG_{ij}\rho ^i_\tau \partial _\tau \chi ^j + \rho
^i_\tau
\hat{M}_{ij} \chi ^j) \bigr\}dtd\tau
}
The mass operator $\hat{M}_{ij}$ here is given by
\eqn\mass{
\hat{M}_{ij}=J_{ij}\partial _t+(D_kJ_{ij})J^{km}\Phi _{m\tau}+
D_i\Phi _{j\tau}
}
Following the standard argument in the quantum field theory,  we want
that
$\hat{M}_{ij}$ be Hermitian. It is necessary for conservation of the
fermionic
charge
\eqn\fecha{
\mu =\oint G_{ij}\rho ^i_\tau \chi ^j\;dt
}
Hermeticity of $\hat{M}_{ij}$ requires
\eqn\symm{
D_{[i}\Phi _{j]\tau}=\partial _{[i}\Phi _{j]\tau}=0
}
If we consider $\Phi _\tau$ as a 1-form on $M$, then \symm is
equivalent to
condition $d\Phi _\tau =0$. As $M$ is a simply connected manifold,
any closed
1-form on it is exact, so there is a {\it function} $H(X,t)$ such
that
\eqn\exha{
\Phi _\tau =-dH(X,t)
}
Thus $\Phi _\tau$ should be a hamiltonian vector field, as we claimed
in
section 4.1.

Note that the mass operator \mass is just a Hessian of the Floer
functional
$S[z]$:
\eqn\massH{
\hat{M}_{ij}={{\delta ^2 S[z]}\over {\delta X^i \delta X^j}}
}
{}From the usual interpretation in the quantum field theory, we
conclude that
the modes having positive masses (eigenvalues of $M_{ij}$),
correspond to the
particles and those having negative masses correspond to
antiparticles\foot{ In
the language of the Morse theory, the particles correspond to the
stable and
antiparticles to the unstable cells of the minimum $z$.}. In order to
have the
stable {\it fermionic} vacuum at each minimum $z\in {\cal Z}$ of $E$,
the Dirac
sea of antiparticles should be completely filled. The different
minima $x,y\in
{\cal Z}$ have the different mass matrices $M_{ij}$ and hence the
different
fermionic vacua. Their fermionic numbers differ by $\mu (x)-\mu (y)$
modulo the
anomaly lattice $2\Gamma$ ( generated by evaluation of $2c_1(M)$ on
the group
$\pi _2(M)$).

It should also be possible to say this other way
\ref\CoJoSe{R.~L.~Cohen,
J.~D.~S.~Jones, G.~B.~Segal {\it Floer's Infinite Dimensional Morse
Theory and
Homotopy Theory} preprint 1993}, \ref\FeFr{B.~Feigin, E.~Frenkel {\it
Affine
Kac-Moody algebras and semi-infinite flag manifolds}
Comm.~Math.~Phys.~ {\bf
128} (1990) 161-185}, using semi-infinite differential forms on the
loop space
${\cal L}M$. The Hessian \massH defines a polarization of the tangent
bundle
$T{\cal L}M$ at the critical point
\eqn\polar{
T{\cal L}M=T{\cal L}M_-\oplus T{\cal L}M_+
}
In turn, it gives a polarization in the Clifford algebra $Cliff$ of
$T{\cal
L}M$ ($Cliff$ is generated by taking the canonical anticommutators
for the
fields $\chi ^i$ and $\rho _\tau ^i$). We may consider a Verma module
of
$Cliff$ associated with this polarization and formally present its
vacuum
vector as $det\;T{\cal L}M_-$ --- a "semi-infinite form". For the
different
critical points in $\cal Z$, the corresponding polarizations in
$Cliff$ are
hopefully compatible with each other and it is possible to define (
modulo
$2\Gamma$ ) a relative degree $\mu (x)-\mu (y)$ of two different
vacua $x$ and
$y$.

To find the physical states \physt we should compute the BRST
cohomology on the
space of the wavefunctionals localized to $\cal Z$. The action of the
BRST
operator $Q$ on their space is encoded in the matrix elements
$<x|Q|y>$ where
$x,y\in {\cal Z}$. We use the path integral representation
\eqn\matbrs{
\langle x|Q|y\rangle =\int  {\cal D}X^i{\cal D}\chi ^i{\cal D}\rho
^i_\alpha \;
Qe^{A[X,\chi
,\rho ]}
}
where the path integral is computed with the boundary conditions
\eqn\bouncon{
\matrix{&X(\tau=-\infty ,t)=x(t)\cr &X(\tau=+\infty ,t)=y(t)\cr}
}
and localize it to the pseudoholomorphic instantons. The term
multiplying the
exponent is the BRST charge
\eqn\brfor{
Q=\oint \bigl(J_{ij}\{\partial _\tau X^i+J^i_{\ k}\partial _tX^k+\Phi
^i_\tau
\}\chi ^j + {1\over 2}D_kJ_{ij}\rho ^i_\tau \chi ^k\chi ^j\bigr)dt\;
}
with fermionic number 1, so \matbrs is zero unless the space of the
instantons
connecting $x$ to $y$ is one dimensional modulo $2\Gamma$. It means
that $\mu
(x)-\mu (y)\equiv 1$ and that \matbrs is localized to the sum over
the same
instantons as appear in the expression \cobou for the coboundary
$\delta$ of
the Floer complex. For each such instanton the $Q$-nontrivial piece
of the
action $A$ gives a factor $\rho =t_1^{s_1}\cdots t_n^{s_n}$ exactly
the same as
the multiplier of the local system $\cal K$. The integration over
fermions
brings the factor $\sigma =\pm 1$  the same as in \cobou. We see that
the
matrix elements \matbrs of the BRST operator $Q$ coincide with that
\cobou of
the coboundary operator $\delta$ of the Floer complex. Thus the Floer
cohomology $H^*_F$ computes just the physical states of our
topological sigma
model.

Actually, we can find the matrix elements of any observable ${\cal
O}_\omega$
in the same fashion:
\eqn\matob{
\langle x|{\cal O}_\omega |y\rangle =\int  {\cal D}X^i{\cal D}\chi
^i{\cal
D}\rho ^i_\alpha
\; {\cal O}_\omega e^{A[X,\chi ,\rho ]}
}
computing the path integral with the same boundary conditions
\bouncon and
localizing it to the instanton configurations. If $\omega$ has a
fermion number
$p$, then \matob is zero unless the space of instantons connecting
$x$ to $y$
is $p$-dimensional, so $\mu (x)-\mu (y)=p$. Repeating the computation
for the
matrix elements of $Q$ we see that \matob coincides with the matrix
element
\wein of the operator $\omega \cup$ in the Floer theory.

Now we can understand better the remark following the formula \wein.
Unless
$Q|x\rangle =Q|y\rangle =0$, i.~e. unless $|x\rangle $ and $|y\rangle
$ are the
physical states, the matrix
element
$\langle x|{\cal O}_{d\Omega} |y\rangle \neq 0$
in general, so \matob depends on the choice of the representative
${\cal
O}_\omega $ for the BRST cohomology class or equivalently, on  the
choice of
the representative $\omega$ for de Rham cohomology $H_{deRham}^*(M)$.
Only
after we restrict to the physical states \physt that the the matrix
elements of
the observables become the functions on $H_{deRham}^*(M)$.

So far we dealt with operators and states independently. But it is a
general
fact that in the topological theories there is a one to one
correspondence
between the operators and states. Now we want to work out this
correspondence
explicitly. It will enable us to identify {\it the matrix elements}
\matob with
the {\it 3-point correlation functions} in the A-model and thereby to
establish
the isomorphism of the Floer's and quantum cohomology.

To do that, we specify to the hamiltonians $H(X)$ independent of $t$
and such
that the corresponding Hamiltonian flows on $M$  have no periodic
trajectories at all\foot{ It is always possible to choose a pair
$J^i_{\ j}$,
$H$ such that this condition is met, see \FI.}. For such $H(X)$, the
critical
loops of the Floer functional $S[z]$ are just the points and coincide
with the
critical points of $H(X)$ on $M$ and the critical set $\cal Z$ is
described by
the usual Morse theory. In \WiIII, which we try to generalize in this
paper,
the Morse theory on $M$ is related to  Supersymmetric Quantum
Mechanics (SQM). This SQM is nothing but the zero-modes approximation
of the
string theory, described by our topological sigma model.

Let us show that the SQM approximation for the matrix elements of the
BRST
operator $Q$ is exact. To see it is true it suffices to show that the
instantons, which appear in \cobou, \matbrs\ are point-like,
i.~e.~correspond
to the propagation of the string as  a point, not as a loop. It would
mean
that only the zero modes are important. But this follows from the
fact that the
relevant in \cobou instantons are {\it isolated}, i.~e.~belong to the
one
dimensional cell of the Floer complex. If an instanton could be
represented as
a joint of a point-like trajectory with a 2-sphere, there would be
2-parametric
freedom to move this sphere around, so such instanton would belong to
at least
2-dimensional cell. This proves that the matrix elements of the
coboundary
operator of the Floer complex coincide with that of the coboundary
operator of
the Witten complex on $M$, so their cohomology, as abelian groups,
are
canonically isomorphic. This is the statement of the Theorem 5 of
\FI, and our
dimension-counting argument is borrowed from its proof.

Of course, the matrix elements \matob of the observables ${\cal
O}_\omega$ are
localized to the 2- and higher dimensional cells of the Floer complex
and
cannot be computed just by SQM.

We are ready now to construct a canonical isomorphism between the
states
$H_F^*$ and the operators $H^*(M)$ of the topological sigma model.
Let us define two nondegenerate pairings of $K$-modules ($K$ is a
Novikov ring)
\eqn\pair{\eqalign{
&(.\;,.\;):\;H_F^p\otimes _KH^p(M)\rightarrow K\cr
&\{.\;,.\;\}:\;H_F^{n-p}\otimes _KH^p(M)\rightarrow K\cr
}}
in a following way. The lowest- and highest degree cohomology
$H^0_W\subset
H^0_F$ and $H^n_W\subset H^n_F$ of the Witten complex on $M$ are
always
generated by one element each. Let us call them $bot$ and $top$
respectively.
They are the $in$ and $out$ vacua of the theory dual to each other:
\eqn\duvac{
\langle bot|{\cal O}_{\Omega}|top\rangle =1
}
where $\Omega$ is the top class in $H^n(M)$
and the correlation functions \FIT of the sigma model are
\eqn\corrf{
\langle {\cal O}_1\cdots {\cal O}_m\rangle =\langle bot|{\cal
O}_1\cdots {\cal
O}_m|top\rangle
}
The formula \corrf is the fundamental relation in the quantum field
theory.
Then the pairings we want to define are:
\eqn\padefI{
(\omega ,x)=\langle bot|{\cal O}_\omega |x\rangle
}
where $\omega \in H^p(M)$, $x\in H^p_F$ and
\eqn\padefII{
\{\omega ,y\}=\langle y|{\cal O}_\omega |top\rangle
}
where $\omega \in H^p(M)$, $y\in H^{n-p}_F$.
Both these pairings are nondegenerate, because in the SQM
approximation, when
$t_i=0$ for all $i$, the pairing $(.\;,\;)$ is the Poincare duality
and
$\{.\;,\;\}$ is the Poincare isomorphism (using the canonical duality
$(H_p(M))^*=H^p(M)$). Hence their determinants, as functions of
$t_1,\ldots
,t_m$, are not equal to zero.

These pairings are related to the two-point correlation functions of
observables (the quantum intersection numbers)
\eqn\rela{
\langle {\cal O}_{\omega _1}{\cal O}_{\omega _2}\rangle =\sum_{x\in
H_F^*}
\langle bot|{\cal
O}_{\omega _1}|x\rangle \langle x|{\cal O}_{\omega _2}|top\rangle
=\sum_{x\in
H_F^*}({\cal
O}_{\omega _1},x)\{{\cal O}_{\omega _2},x\}
}
which is just a statement of completeness of the physical states of
the quantum
theory. Thus the pairings above give both the isomorphism between the
operators
and the states:
$h^p:\; H^p_F=H^p(M)$
{\it and} the quantum intersection matrix. The 3-point functions are
related to
the matrix elements \matob by the formula
\eqn\finrel{
\langle {\cal O}_{\omega _1}{\cal O}_{\omega _2}{\cal O}_{\omega
_3}\rangle
=\sum_{x,y\in
H_F^*}({\cal O}_{\omega _1},x)\langle x|{\cal O}_{\omega _2}|y\rangle
\{{\cal
O}_{\omega _3}
,y\}
}
Note that the (super)commutativity and associativity of the algebra
of
observables ${\cal O}_\omega$ give the relations for the matrix
elements
\matob, equivalent to the supercommutativity and associativity of the
cup
operation $\omega \cup$ in the Floer theory.

{\it In other words, the cohomology $H^*(M)$ with the cup product
multiplication from the Floer theory is isomorphic to the quantum
cohomology
ring.}

\newsec{Some examples}
Now we would like to present some computations, partly described in
\FI, showing how our "matrix" approach really works.
\subsec{Projective spaces}
This is the simplest possible example. The homology $H^*_F$ and the
action of
$H^*_{deRham}({\bf CP^1})$ on it were computed by Floer himself in
\FI. It is
very  instructive to repeat his argument.

Let us take the hamiltonian vector field
\eqn\vefi{
v=z{\partial \over {\partial z}} + \bar{z}{\partial \over {\partial
\bar{z}}}
}
It does not have periodic trajectories and only fixed points are the
north (N)
and the south (S) poles of the sphere $S^2={\bf CP^1}$. Thus in the
loop space
${\cal L}{\bf CP^1}$ the critical set of the Floer functional \Morf
consists
just of these two points. By the universal cover map $\widehat{{\cal
L}{\bf
CP^1}}\longrightarrow {\cal L}{\bf CP^1}$ the set $(N,S)$ is covered
by the
points
$(\ldots ,N^{(-1)},S^{(-1)},N^{(0)},S^{(0)},N^{(1)},S^{( 1)},\ldots
)$ so that
$N^{(k)}\longrightarrow N$ and $S^{(k)}\longrightarrow S$. In fact,
this
picture represents \FeFr the affine Weyl diagram for
$\widehat{sl(2)}$.

The trajectories going from $N^{(k)}$ to $S^{(k)}$ are the
"classical" ones.
Downstairs  they cover a 2-parametric set of trajectories of the
vector field
$v$. As a set of points, this set coincides with the base projective
line
itself.

On the other hand, the trajectories from $S^{(k)}$ to $N^{(k+1)}$ are
essentially stringy --- they have homotopic type of ${\bf CP^1}$.
Again, there
is a 2-parametric family of them due to the action of ${\bf C^*}$ on
the world
sheet (a sphere with two marked points). Each such trajectory covers
the base
${\bf CP^1}$.

Again, this can be represented by the affine Weyl diagram, if we
denote the
2-dimensional cells of the Floer complex by the arrows connecting the
vertices
\break
$(\ldots , N^{(-1)}, S^{(-1)}, N^{(0)}, S^{(0)}, N^{(1)}, S^{( 1)},
\ldots )$.

As in the usual Morse theory for ${\bf CP^1}$, there is no
1-dimensional cells,
hence the coboundary operator is trivial and the Floer cohomology is
represented by $N$ (of degree 0) and $S$ (of degree 2). Integrated
over ${\bf
CP^1}$, the first Chern class of the tangent bundle gives 2, so the
fermionic
number anomaly is 4.

Now let us find the matrix elements of the generator $x$ of
$H^*_{deRham}({\bf
CP^1})$ between $S$ and $N$. The element $\langle N|x|S\rangle $
comes from
integration of
(the pullback of) $x$ on ${\cal C^2}(N,S)={\bf CP^1}$, each
trajectory is
homotopicaly trivial in the loop space. Therefore, the "classical"
answer
$\langle N|x|S\rangle =1$ holds true.

Unlike the usual Morse theory, there is also a nontrivial matrix
element
$\langle S|x|N\rangle $.
Formally it is possible because $\mu (N)-\mu (S)=2-4=-2\equiv 2 (mod
4)$. There
is indeed a 2-parametric family of stringy trajectories from $S$ to
$N$ as we
saw above. The integral of the pullback of $x$ over it is again the
integral of
$x$ over ${\bf CP^1}$. But now the homotopic type of each trajectory
is not
trivial and to obtain the matrix element \wein we need to multiply
the integral
of $x$ by \weght equal to $t=\exp{-\int k}$. Thus $\langle
S|x|N\rangle =t$.
Therefore, the matrix representation of the operator ${\cal O}_x$ is
\eqn\marep{
{\cal O}_x = \pmatrix{ &0 &1 \cr &t &0}
}
It satisfies the relation
\eqn\sirin{
{\cal O}_x {\cal O}_x =t
}
well known for ${\bf CP^1}$ sigma model.

A straightforward generalization of the example above is ${\bf
CP^n}$. It was
also done in \FI. Again, there is no 1-dimensional cells in the Floer
complex
and $H_F^*$ is spanned by the critical points of ($S^1$ equivariant)
Morse
function on ${\bf CP^n}$ of degrees $0,2,\ldots ,2n$. The fermionic
number
anomaly is $2(n+1)$.

The de Rham cohomology $H^*_{deRham}({\bf CP^n})$ is generated as a
ring by
one element $x$. We can again find the matrix representation for the
operator
${\cal O}_x$, which generates the {\it quantum} ring. In this example
it is
also true that ${\cal O}_{x^k}{\cal O}_{x^l}={\cal O}_{x^{k+l}}$, for
$k+l<n+1$; nothing like this should be true in general.

There are two simple remarks we want to make before we write the
formula for
${\cal
O}_x$. Let $M$ be an almost K\"ahler manifold with $c_1 > 1$ and
$x\in H^2(M)$.
Then the matrix elements of $x$ on  the main diagonal and above, like
$\langle z_m|x|z_{m'}\rangle $, $m'\geq m$, do not depend on
$t_1,\ldots ,t_n$
and can be
computed classically.
Indeed, to get the matrix element the pullback of $x$ should be
integrated over
some two-cycle of the Floer complex. But if this cycle consisted of
stringy
trajectories, (which would lead to $t_i$ dependence), then by the
index theorem
it would belong to the cell of the dimension at least 4.
The other simplification for the matrix elements of $x\in H^2(M)$ is
that the
2-cells of the Floer complex, for the purposes of intersection
theory, are
representable by the 2-cells on $M$ itself, just like it was for
${\bf CP^1}$.
This is because these 2-cells consist either of point-like classical
trajectories or of a single 2-sphere in $M$, reparametrized by ${\bf
C^*}$.

All the matrix elements below the main diagonal are always due to the
stringy
paths.
The formula for ${\cal O}_x$ is
\eqn\cpn{
{\cal O}_x= \pmatrix{&0 &1 &0 &\cdots &0 \cr
                     &0 &0 &1 &\cdots &0 \cr
                     &\cdot &\cdot &\cdot &\cdots &\cdot \cr
                     &0 &0 &\cdots &0 &1 \cr
                     &t &0 &\cdots &0 &0}
}
To obtain \cpn we note that as $c_1 = (n+1)x$, only $<z_{2n}|x|z_0>$
can be a
nontrivial matrix element below diagonal (as $0-2n+2(n+1)\cdot 1=2$).
The
2-cell ${\cal C}^2(z_{2n},z_{0})$ consists of paths from $z_{2n}$ to
$z_0$ of
degree 1 and coincides with a straight line $(z_{2n},z_{0}$ for the
appropriate
choice of hamiltonian. It explains the degree 1 of $t$ as well as a
numerical
coefficient 1 before it in \cpn. The operator ${\cal O}_x$ satisfies
\eqn\cpnn{
{\cal O}_x^n=t
}

\subsec{Flag spaces}
A less trivial generalization of the example with ${\bf CP^1}$ is the
flag
space
$Fl_n$, which can be realized as a coset
\eqn\fl{
Fl_n=U(n)/(U(1))^n
}
Its second cohomology group is ${\bf Z^{n-1}}$ generated by
$x_1,\ldots,x_{n-1}$. Cohomology of $Fl_n$, as a
ring\ref\BoTu{R.~Bott, F.~Tu
{\it Differential Forms in Algebraic Topology}}, is generated by
$x_i$ with the
relations which are homogenious components of the single relation
\eqn\flrel{
\Pi _{i=1,\ldots ,n}(1+x_i)=1
}

We will compute the matrix elements of $x_1, x_2$ in the Floer theory
for the
simplest nontrivial example of 3-dimensional flag space $Fl_3$. A way
we do
this, using an affine Weyl group for $\widehat{sl(3)}$, can be
generalized for
all other flag spaces.

The Floer theory for the loop spaces of flags was considered in
\FeFr. The
critical point set in ${\cal L}Fl_n$ is parameterized by the elements
of the
{\it finite} Weyl group. The cover map $\widehat{{\cal
L}Fl_n}\longrightarrow
{\cal L}Fl_n$ covers it by the {\it affine} Weyl group. A simplest
example to
look at is $Fl_2={\bf CP^1}$, considered in the previous section.
There are no
1-cells in the Floer complex; to describe all the 2-cells we need an
auxiliary
geometric construction.

Let us think of the elements $x_1, x_2$ as of two simple roots of
${sl(3)}$.
There are two standard embeddings of $sl(2)$ to $sl(3)$ sending a
simple root
of $sl(2)$ either to $x_1$ or to $x_2$. These maps can be continued
to the maps
of groups $SL(2)\longrightarrow SL(3)$ which, in turn, induce two
maps  ${\bf
CP^1}\longrightarrow Fl_3$ which send the generator $x$ of $H_2({\bf
CP^1})$ to
either $x_1$ or $x_2$.

This construction extends to give the maps of the (universal covers
of) {\it
loop} spaces. Looking at the diagram for the affine Weyl group for
$\widehat{sl(3)}$\foot{It is a 2-lattice with hexagonal point group.}
we see
that it can be covered by the straight lines parallel to any simple
root. Each
such straight line correspond to some particular embedding
$\widehat{{\cal
L}{\bf CP^1}}\longrightarrow \widehat{{\cal L}Fl_3}$. A pattern of
critical
points along the straight line coincides with that one for
$\widehat{{\cal
L}{\bf CP^1}}$: when projected downstairs to the base flag space
$Fl_3$ it
covers {\it two} (which ones, depends on the particular straight
line) critical
points we denote by $N$, $S$. These two points we can identify with
the
critical points on the preimage ${\cal L}{\bf CP^1}$. The arrangement
of
critical points upstairs is
\break
$(\ldots , N^{(-1)}, S^{(-1)}, N^{(0)}, S^{(0)}, N^{(1)}, S^{( 1)},
\ldots )$.

Now let us choose on $Fl_3$ such hamiltonian vector field that it
respects both
embeddings ${\bf CP^1}\longrightarrow Fl_3$. In the
finite-dimensional Morse
theory it means that the {\it whole flow}, connecting $N$ to $S$,
belongs to
the preimage ${\bf CP^1}$. Then this property is promoted to the loop
spaces.
It means that the matrix elements of $x_1, x_2$ between $N$ and $S$
can
effectively be computed within a Floer theory for a projective line
${\bf
CP^1}$. Not every pair of critical points whose indices differ by
2$(mod\ 4)$
lies on a straight line parallel to a simple root. For them, the
degree count
shows that only the matrix elements $\langle top|x_i|bot\rangle
\propto t_1
t_2$ can be
non-zero, and the commutativity of two matrix operators ${\cal
O}_{x_1}$,
${\cal O}_{x_2}$ fixes the integral coefficients in front of $t_1
t_2$.

Ultimately, the operators ${\cal O}_{x_1}$ and ${\cal O}_{x_2}$ are
given by
\eqn\maxI{
{\cal O}_{x_1}=
\pmatrix{  &0  &1  &0  &0  &0  &0  \cr
           &t_1 &0 &0 &1 &0 &0  \cr
           &0 &0 &0 &-1 &-1 &0  \cr
           &0 &0 &-t_1 &0 &0 &0 \cr
           &0 &0 &0 &0 &0 &1    \cr
           &t_1 t_2 &0 &0 &0 &t_1 &0 }
}
\eqn\maxII{
{\cal O}_{x_2}=\pmatrix{  &0 &0 &1 &0 &0 &0   \cr
                          &0 &0 &0 &-1 &-1 &0 \cr
                          &t_2 &0 &0 &0 &1 &0 \cr
                          &0 &0 &0 &0 &0 &-1  \cr
                          &0 &-t_2 &0 &0 &0 &0 \cr
                          &-t_1 t_2 &0 &0 &-t_2 &0 &0 }
}
They generate an algebra with relations
\eqn\Relns{\eqalign{
&{\cal O}_{x_1}^2 + {\cal O}_{x_1}{\cal O}_{x_2} + {\cal
O}_{x_2}^2=(t_1+t_2)
\cr
&{\cal O}_{x_1}^3 = t_1 {\cal O}_{x_1} - t_1 {\cal O}_{x_1} \cr
&{\cal O}_{x_2}^3 = t_2 {\cal O}_{x_2} - t_2 {\cal O}_{x_1} \cr
&{\cal O}_{x_1}^2 {\cal O}_{x_2} + {\cal O}_{x_1} {\cal O}_{x_2}^2 =
t_2 {\cal
O}_{x_1} + t_1 {\cal O}_{x_2}
}}
which can be considered as a two-parametric deformation of \flrel.

\subsec{Calabi-Yau manifolds}
Almost all the known nontrivial examples of quantum cohomology of CY
manifolds
are obtained using mirror symmetry. Mirror symmetry reduces the
problem of
computation of quantum cohomology $H_Q^*(M)$ to some computation in
the
Variation of Mixed Hodge Structure Theory of the mirror pair $W$. The
latter
problem is usually relatively easier.

As a first application of the general theory we may just transfer all
these
results to the Floer theory, which give the first examples of the
latter for
the CY manifolds.
Note, that unless we consider the Floer complex over the Novikov
ring, the
boundary operator is not defined --- its matrix elements are
represented by the
divergent series due to summation over all maps of the cylinder to
itself. The
same is true for the matrix elements of de Rham cohomology.

On the other hand, in the string theory computations for the
3-dimensional CY,
we are mostly interested in the matrix elements of the second de Rham
cohomology ( $H^{1,1}(M)$, to be precise), which generate the
marginal
deformations of the corresponding superconformal theory. A remark in
the
section about projective spaces tells that these can be computed just
looking
at the 2-cycles of the CY manifold itself, without going up to the
loop space.
We hope it may make possible to do a direct computation in some
examples.
\newsec{Conclusion: Noncompactness and non generic moduli of
instantons.}
Roughly speaking, there are two ingredients in the Floer theory. One
of them,
the algebraic one, is very neat, we tried to stress it in the section
3. The
second --- analytic --- ingredient is what makes the Floer theory so
difficult.
In this paper we tried to substitute the language of the topological
quantum
field theory for the analytic language of Floer {\it et. al.} From
the
mathematician's point of view, it may seem as trading a bad thing for
the worse
one.

A real advantage of our approach  is that it makes the
independence of $H_F^*$ of the hamiltonian $H$ and the almost complex
structure
$J$ an
immediate consequence of the BRST invariance, {\it without any
assumptions
about} $H$. In particular, the choice $H=0$ is admissible. On the
other hand,
for $H=0$ and $J$ being a true complex structure the path integral
defining the
sigma model has a combinatorial
definition \WiII\ in terms of the moduli spaces of holomorphic
instantons. It
is
an object of study of algebraic geometry which does not require
infinite-dimensional analysis to deal with.

We hope that  it  may be helpful to think about the same object using
two
languages.
For the topological theory, it may also be useful to have a picture
where all
the
physical states are localized to the particular loops.

There are two fine points which we want to tuch upon in conclusion.
The first
is a problem
 of non-compactness of moduli space of  (almost) holomorphic
instantons. The
effects on the ``infinity" can be crucial when considering the
effects of
``BRST trivial" deformations.  Landau-Ginzburg  $A_n$ model with
superpotential
$W=x^{n+2}$ is an example. The deformation $W\longrightarrow
W+\epsilon
x^{n+3}$ is ``BRST trivial" but changes the theory completely:
$A_n\longrightarrow A_{n+1}$.  It is important to realize that the
problem in
this example appears  already at the level of BRST cohomology:  the
deformation
brings  ``from    infinity" one extra  ground state or equivalently,
makes
physical an extra  operator $x^{n+1}$.  This changes completely the
structure
of  the operator state correspondence. In particular, the dependence
of  the
vacuum $|top\rangle$ corresponding to the unity operator  is {\it
not}
continuous.  This makes unapplicable the standard perturbative
argument used to
prove the BRST invariance of the correlation functions.  And indeed,
the
correlation functions are not even continuous at  $\epsilon =0$.

Let us examine in light of that example the situation occuring in our
theory
with $H\neq 0$.  We have seen in Section 4 that the set of local
physical
operators (BRST cohomology) is {\it independent} of  $H$  (even at
the level of
representatives).   We need to prove that the correlation functions
of these
operators are also independent of $H$.  The correlation function can
be
considered as the matrix element  $\langle {\cal O}_1\cdots {\cal
O}_m\rangle
=\langle  bot|{\cal O}_1\cdots {\cal O}_m|top\rangle$ of the product
${\cal
O}_1\cdots {\cal O}_m$ between two vacua.   By the state operator
correspondence the vacuum $|bot \rangle$ is mapped to 1 and the
vacuum  $|top
\rangle$ is mapped to ${\cal O}_\Omega$ where $\Omega$ is a top form
on $M$.
They are normalized by
\eqn\normal{
\langle bot|{\cal O}_\Omega |top\rangle=1
}
{\it If} the H deformation  changes the vacua  $|top \rangle$ and/or
$|bot
\rangle$ discontinuously the perturbation theory argument cannot be
applied and
  the correlation functions may change. Let us show it does not
happen. This is
obvious when $M$ is  Calabi Yau so  that the sigma model is
conformal. Then the
fermionic number is strictly conserved. Since $|bot \rangle$ is the
{\it
unique} state with fermionic number 0 and  $|top \rangle$ is the {\it
unique}
state with fermionic number $n=dim\, M$, any perturbation which
preserves BRST
cohomology can only {\it rescale} them.  But the relevant scaling
factor is
fixed by \normal.  This proves the statement.

If  $M$ is not Calabi Yau,  the fermionic number is  conserved only
modulo
fermionic anomaly.  Still we can define   $|top \rangle$ and  $|bot
\rangle$
using the operator state correspondence as above (one can also use
the SQM
approximation). Suppose that  the  deformation could lead to
\eqn\deform{\eqalign{|top \rangle \longrightarrow |top' \rangle
&=|top \rangle
+\sum r_p\,q_1^{-p_1}\ldots q_k^{-p_k}|p \rangle,\cr
|bot \rangle\longrightarrow  |bot' \rangle &=|bot\rangle +\sum
s_{p'}\,q_1^{N_1-p'_1}\ldots q_k^{N_k-p'_k}|p' \rangle.
}}
  In \deform,  $|p \rangle$ and $|p' \rangle$ are the ground states
of (multi)
degrees $(p_1,\ldots, p_k)$ and $(p'_1,\ldots, p'_k)$.  $(N_1,\ldots,
N_k)$ is
a multidegree of  $|top\rangle$. The coefficients $r_p$ and $s_{p'}$
are
$q$-independent. The formula \deform\ gives the most general form of
the
deformation compatible with anomalous conservation of fermionic
number. Having
the $q$-dependence  completely fixed, let us consider the
quasiclassical limit
$q_i \longrightarrow 0$.  Since the fermionic number is conserved in
this
limit, we expect that the deformation should tend to zero with $q_i$.
We see
this is true for  $|top\rangle$ for any $s_{p'}$. On the other hand,
it can
only be true for $|bot\rangle$ if all the coefficients $r_p$ are
equal  to
zero.

Now let us compute the matrix element  $\langle bot'|{\cal
O}_\omega|top'\rangle$ of  any operator ${\cal O}_\omega$, $deg\,
\omega <n$.
On one hand, since $|bot'\rangle=|bot\rangle$, this matrix element
is given by
$$\langle bot'|{\cal O}_\omega|top'\rangle =\langle bot|{\cal
O}_\omega|top\rangle +\sum s_{p'}\,q_1^{N_1-p'_1}\ldots
q_k^{N_k-p'_k}\langle
bot|{\cal O}_\omega|p' \rangle.$$
On the other hand, one can compute it directly.  Since $deg\, {\cal
O}_\omega<deg\,|top \rangle -deg\,|bot \rangle =n$ and $c_1(M)$ is
positive,
this matrix element does not have instanton corrections, can be
computed in SQM
and equals zero. Of course, the same is true about
 $\langle bot|{\cal O}_\omega|top\rangle$ so we end up with equation
$(\omega ,\sum [s_{p'}\,q_1^{N_1-p'_1}\ldots q_k^{N_k-p'_k}] p') =0 $
for any
$\omega \in H^*(M)$, where we denoted $(\omega,p')=\langle bot|{\cal
O}_\omega|p' \rangle$. It was explained in Section 4 that the pairing
$(.\,
,.\,)$ is nondegenerate over ${\bf C}[q_i,q_i^{-1}]$. Therefore, the
only
solution is $s_{p'}=0$.

We just have derived that  the vacua  $|top \rangle$ and $|bot
\rangle$  cannot
be deformed by any perturbation not changing the BRST cohomology.
The $H$
independence of the correlation functions follows.  Indeed, since
the vacua
remain the same  we can treat the  $H$ deformation within the
perturbation
theory:
\eqn\pert{\eqalign{
\langle {\cal O}_1\ldots {\cal O}_k\rangle _{\varepsilon H}=& \langle
bot|
{\cal O}_1\ldots {\cal O}_k\, e^{\varepsilon \{Q,A\} } |top \rangle =
\langle
bot| {\cal O}_1\ldots {\cal O}_k\, (1+\sum_{n=1}^{\infty}
{\varepsilon ^n\over
n!} \{Q,A\}^n ) |top \rangle =\cr
=&\langle bot| {\cal O}_1\ldots {\cal O}_k\, 1\, |top \rangle=\langle
{\cal
O}_1\ldots {\cal O}_k\rangle _{0}.
}}

The second point we want to discuss is the problem of  ``non generic"
moduli
spaces of instantons.  It might happen (and almost always happens in
the
(almost) holomorphic case) that the dimension of the some connected
component
of the moduli space of instantons is
strictly greater than its lower bound given by the Riemann-Roch.
(Then in order
to obtain the correlation function one has to compute the Euler
characteristic
of a certain vector
bundle on this (properly compactified) component of the moduli space
\WiII,
\AsMo.)  From the point of view of  the real differential  geometry,
this
situation is not generic.  The key point here is that there are no
``generic
deformations" in the algebraic geometry in a sense of differential
geometry, so we cannot resolve the ``degeneration" within (almost)
complex-analytic
setup.

The way out is to break the holomorphicity. This is exactly what we
did with
\pshol. Deformed by $H$, the theory is no longer complex-analytic.
(For
example,
the real dimension of cells of the Floer complex can be odd and the
intersection numbers can have arbitrary signs.) Thus the H
deformations enable
one to reduce
any degenerate situation to the generic one. The topological
correlation
functions being $H$ independent, the  results of
computation are not affected.  (That is why we do not have to discuss
the
multiplication in Floer's theory when the situation is not generic.
In
principle, it could be done along the lines of
\WiII, \AsMo ).

\bigbreak\bigskip\bigskip\centerline{{\bf Acknowledgements}}\nobreak
I thank A.~Astashkevich, M.~Bershadsky, S.~P.~Novikov, S.~Piunikhin
and C.~Vafa
for the useful discussions. The remarks of D.~Kazhdan and I.~Singer
were
important at the  final stage of the work.

Research supported in part by the Packard Foundation and by NSF grant
PHY-87-14654

\listrefs
\bye